\documentclass[12pt]{article}
\usepackage{epsfig,amssymb,euscript}
\usepackage{amsmath}
\newcommand{\RR}{\mathbb{R}}
\newcommand{\CC}{\mathbb{C}}

\def\be#1\ee{\begin{equation}#1\end{equation}}
\newcommand{\bea}{\begin{eqnarray}}
\newcommand{\eea}{\end{eqnarray}}
\newcommand{\ba}{\begin{array}}
\newcommand{\ea}{\end{array}}

\newcommand{\p}[1]{(\ref{#1})}

\def\bbox{{\,\lower0.9pt\vbox{\hrule \hbox{\vrule height 0.2 cm
\hskip 0.2 cm \vrule height 0.2 cm}\hrule}\,}}
\newcommand{\dsl}{\pa \kern-0.5em /}

\newcommand{\nn}{\nonumber \\}

\DeclareMathOperator{\Hol}{Hol}

\DeclareMathOperator{\Tr}{Tr}
\DeclareMathOperator{\im}{Im}

\newcommand{\rep}[1]{\mathbf{#1}}

\newcommand{\e}{\epsilon}

\def\a{{\alpha}}
\def\s{{\sigma}}
\def\b{{\beta}}
\def\l{\lambda}

\def\ds{\raise.15ex\hbox{/}\kern-.57em\partial}
\def\Ds{\,\raise.15ex\hbox{/}\mkern-13.5mu D}
\begin{document}

%Section numbering????
\makeatletter
\renewcommand{\theequation}{\thesection.\arabic{equation}}
\@addtoreset{equation}{section} \makeatother

\baselineskip 18pt

%%%%%%%%%%%%%%%% title page %%%%%%%%%%%%%%%%%%%%%%%%%%%%%%%%%%%%

\begin{titlepage}
\vfill
\begin{flushright}
QMUL-PH-02-09\\
hep-th/0205050\\
\end{flushright}

\vfill

\begin{center}
\baselineskip=16pt
{\Large\bf $G$-Structures and Wrapped NS5-Branes}
\vskip 10.mm {Jerome P. Gauntlett$^{*\dagger1}$, Dario
Martelli$^{*2}$, Stathis Pakis$^{*3}$
and Daniel Waldram$^{*4}$}\\
\vskip 1cm
%\vfill
$^*${\small\it
Department of Physics\\
Queen Mary, University of London\\
Mile End Rd, London E1 4NS, U.K.\\} \vskip 0.5cm \vskip 0.5cm
$^\dagger${\small\it Isaac Newton Institute for Mathematical Sciences\\
University of Cambridge\\
20 Clarkson Road, Cambridge, CB3 0EH, U.K.\\} \vspace{6pt}
\end{center}
\vfill
\par
\begin{center}
{\bf ABSTRACT}
\end{center}
\begin{quote}
We analyse the geometrical structure of supersymmetric solutions of
type~II supergravity of the form $\RR^{1,9-n}\times M_n$ with
non-trivial NS flux and dilaton. Solutions of this type arise
naturally as the near-horizon limits of wrapped NS fivebrane
geometries. We concentrate on the case $d=7$, preserving two or four
supersymmetries, corresponding to branes wrapped on associative or
SLAG three-cycles. Given the existence of Killing spinors, we show that
$M_7$ admits a $G_2$-structure or an $SU(3)$-structure, respectively,
of specific type. We also prove the converse result. We use the
existence of these geometric structures as a new technique to derive
some known and new explicit solutions, as well as a simple theorem
implying that we have vanishing NS three-form and constant dilaton
whenever $M_7$ is compact with no boundary. 
The analysis extends simply to other
type II examples and also to type I supergravity.   
\end{quote}

\vfill 

\hrule width 5.cm \vskip 5mm {\small
\noindent $^1$ E-mail: j.p.gauntlett@qmul.ac.uk \\
\noindent $^2$ E-mail: d.martelli@qmul.ac.uk \\
\noindent $^3$ E-mail: s.pakis@qmul.ac.uk \\
\noindent $^4$ E-mail: d.j.waldram@qmul.ac.uk \\
}
\end{titlepage}
%%%%%%%%%%%%%%%%%%%%%%%%%%%%%%%%%%%%%%%%
\setcounter{equation}{0}

\section{Introduction}

Solutions of type II supergravity corresponding to NS fivebranes
wrapping supersymmetric cycles provide an interesting arena for
studying the holographic duals of supersymmetric Yang-Mills (SYM)
theory \cite{malnuntwo}. Solutions, in the ``near-horizon
limit'', have now been found for a number of different cases
\cite{agk}--\cite{kapustin}. In each case the final geometry is of
the form $\RR^{1,5-d}\times M_{4+d}$, where $d$ is the
dimension of the cycle on which the fivebrane is wrapped, and the NS
three-form $H$ and the dilaton are non-trivial. 

In \cite{GKMW} various aspects of the geometry of such
supersymmetric solutions was elucidated. The key input is the
existence of Killing spinors describing the preserved supersymmetries. 
In type II supergravity, the vanishing of the supersymmetry variation
of the gravitino implies that any Killing spinor is parallel with
respect to one of two connections $\nabla^\pm$ with totally
anti-symmetric torsion $\pm\frac{1}{2}H$. This implies that $M_{4+d}$
admits certain geometric structures and the vanishing of the variation
of the dilatino imposes additional conditions on the structures. It
was shown in \cite{GKMW}, following \cite{GKPW}, that the resulting
structures give rise to generalised calibrations \cite{GPT}.

Here we would like to continue the investigations of~\cite{GKMW},
again assuming only the existence of particular sets of Killing
spinors. Thus while motivated by considering solutions for wrapped NS
five-branes, the results apply generally to supersymmetric backgrounds
with non-trivial NS-flux $H$ and dilaton.  For definiteness we will
focus on seven-dimensional geometries $M_7$, though it is clear that
the analysis generalises to all cases discussed in \cite{GKMW}. Two
distinct types of geometry arise. The first is when the Killing spinors
are all parallel with respect to the same connection, say
$\nabla^+$. Geometrical structures of this type have been discussed
in~\cite{strominger,Hull,IvPap,Ivanov1,Ivanov2,Iv3} as well
as~\cite{GKMW}. The second, new, type \cite{GKMW} 
is when there are some Killing
spinors parallel with respect to $\nabla^+$ and some with respect to
$\nabla^-$. 

For our particular example, seven-dimensional geometries arise when
fivebranes wrap associative three-cycles or SLAG three-cycles . The
geometries are distinguished by the fact that the former is of the
first type with a single Killing spinor parallel with respect to
$\nabla^+$. The latter case, on the other hand, is of the second type,
preserving twice as much supersymmetry, with two Killing spinors
$\epsilon^\pm$, parallel with respect to $\nabla^\pm$ respectively. For
the generic case with a single Killing spinor, the seven-dimensional
geometry admits a $G_2$-structure of a specific type, to be reviewed
below. On the other hand, the seven-dimensional geometries with two
Killing spinors $\epsilon^\pm$ admit two $G_2$-structures, or more
precisely an $SU(3)$ structure, again of a specific type to be
discussed. Note that because of the non-vanishing intrinsic torsion
the $SU(3)$-structure does not imply that the manifold is a direct
product of a six-dimensional geometry with a one-dimensional
geometry. However, there is a product structure which does allow the
metric to be put into a canonical form with a six-one split as we shall
discuss. 

The geometrical structures defined by the preserved
supersymmetries can equivalently be specified by tensor fields
satisfying first-order differential equations. These give a set of 
necessary conditions imposed by preservation of supersymmetry and
the equations of motion. We will show that for the particular cases of
$G_2$- and $SU(3)$-structures mentioned above, these conditions are
also sufficient. From the derivation, it is clear that should always
be possible to find such a set of conditions. Note that, for the cases
of single $G_2$- and $Spin(7)$-structures, this issue was also analysed
in~\cite{Ivanov1,Ivanov2} and~\cite{Iv3} respectively, although
these works did not consider the relationship between Killing
spinors and the equations of motion, as we shall here.  

One of the motivations for this work was to see if these
sufficient conditions just mentioned could provide a new method
for constructing supersymmetric solutions describing wrapped
fivebranes. This would provide an alternative to the ``standard''
two-step construction \cite{malnun,malnuntwo} of first finding a
solution of $D=7$ gauged supergravity and then uplifting to $D=10$. We
shall show that some known solutions can be recovered in this way. In
addition to providing a direct $D=10$ check of the solutions, this makes
their underlying geometry manifest. We will also use the method to
construct a new solution that describes a fivebrane wrapping a
non-compact associative three-cycle. It is a co-homogeneity one solution
with principle orbits given by $SU(3)/U(1)\times U(1)$.

Recall that the solution of~\cite{gr,GKMW} describing fivebranes
wrapping SLAG three-cycles was argued to be dual to pure
$\mathcal{N}=2$, $D=3$ SYM. By analogy with~\cite{agk,maldnast} it
seems likely that there are more general solutions that would be dual
to $\mathcal{N}=2$, $D=3$ SYM with a Chern--Simons term. These seem
difficult to find using gauged supergravity. Unfortunately they also
seem to be difficult to obtain using the methods to be described
here. In particular, to recover the known solution of~\cite{gr,GKMW}
one is first naturally led to partial differential equations, and it
is somewhat miraculous that there is change of coordinates that leads
one to the relatively simple solution obtained in~\cite{gr,GKMW} using
the gauged supergravity approach. 

The type II supergravity solutions for wrapped branes that are
dual to quantum field theories are non-compact. As somewhat of an
aside we also prove a simple vanishing theorem for compact
manifolds. In particular, we show that the expression for the
three-form $H$ in terms of the $G_2$-structure allows one to prove
that on a compact manifold without boundary given $dH=0$, the
three-form must necessarily vanish and the dilaton $\Phi$ is
constant. There are analogous expressions for $H$ in terms of
generalised calibrations for other fivebrane geometries \cite{GKMW}, 
and hence this result generalises easily. 

The plan of the rest of the paper is as follows. We begin in
section~2 with some general discussion of $G$-structures and
$G$-invariant tensors using $G_2$ as our example. In
section~3 we review and extend what is known about the geometry
with $G_2$ structure that arises when type II fivebranes wrap
associative three-cycles. We also prove the vanishing theorem for
compact manifolds. Section~4 discusses the geometry that arises
when fivebranes wrap SLAG three-cycles. In this case there are
two $G_2$ structures or equivalently an $SU(3)$ structure.
Section~5 uses the necessary and sufficient
conditions for $G_2$-structures admitting
Killing spinors as a technique to rederive some known solutions as
well as deriving a new solution that describes a fivebrane wrapping a
non-compact associative three-cycle. In section~6 we use the analogous
conditions for $SU(3)$-structures to derive BPS equations for solutions
corresponding to fivebranes wrapped on SLAG three-cycles. We conclude
in section~7 by discussing how the results would extend to fivebranes
wrapping other supersymmetric cycles as discussed in \cite{GKMW} and
we also briefly comment on the extension to type I supergravity.

\section{$G$-structures}

In this section we review the notion of $G$-structure and
$G$-invariant tensors on a Riemannian manifold $M$ and the relation to
intrinsic torsion and holonomy. Though the discussion is general, our
examples will concentrate on the case relevant here of $G_2$-structure
on a seven-manifold. Further details can be found, for example,
in~\cite{Joyce} and~\cite{sal}. 

In general the existence of a $G$-structure on an $n$-dimensional
Riemannian manifold means that the structure group of the frame bundle
is not completely general but can be reduced to $G\subset O(n)$. Thus,
for $G_2$-structures on a seven-manifold, the structure group
reduces to $G_2\subset SO(7)\subset O(7)$.

An alternative and sometimes more convenient way to define
$G$-structures is via $G$-invariant tensors. A non-vanishing,
globally defined tensor $\eta$ is $G$-invariant if it is invariant
under $G\subset O(n)$ rotations of the orthonormal frame. Since $\eta$
is globally defined, by considering the set of frames for which $\eta$
takes the same fixed form, one can see that the structure group of the
frame bundle must then reduce to $G$ (or a subgroup of $G$). Thus the
existence of $\eta$ implies we have a $G$-structure. Typically, the
converse is also true. Recall that, relative to an orthonormal frame,
tensors of a given type form the vector space, or module, for a
given representation of $O(n)$. If the structure group of the frame
bundle is reduced to $G\subset O(n)$, this module can be decomposed into
irreducible modules of $G$. Typically there will be some type of tensor 
that will have a component in this decomposition which is invariant under $G$. 
The corresponding vector bundle of this component must be trivial, and
thus will admit a globally defined non-vanishing section $\eta$.   

To see how this works in the case of $G_2$-structures, consider the
three-form on $\RR^7$ given by  
\begin{equation}
   \phi_0=dx^{136}+dx^{235}+dx^{145}-dx^{246}-dx^{127}-dx^{347}-dx^{567}
\end{equation}
where $dx^{ijk}=dx^i\wedge dx^j\wedge dx^k$ and let
$g_0=dx_1^2+...+dx_7^2$ denote the standard Euclidean metric. The
group $G_2$ can be defined as the subgroup of the $O(n)$ symmetries of
$g_0$ which leaves $\phi_0$ invariant. A seven-dimensional manifold
$M_7$ then admits a $G_2$-structure if and only if there is a globally
defined three-form $\phi$ on $M_7$ which is $G_2$-invariant. That is,
at each point on $M_7$ we can consistently identify the three-form
$\phi$ with the standard $G_2$-invariant three-form $\phi_0$. Note
that given $\phi_0$ we also have the metric $g_0$, an orientation
$dx^{1\dots7}$ and the Hodge-dual four-form ${*\phi}_0$. Thus given a
$G_2$-invariant $\phi$ on $M_7$ we actually also get an associated
metric $g$ and four-form ${*\phi}$ on $M_7$ such that $(\phi,{*\phi},g)$
are identified under the map to $\RR^7$ with $(\phi_0,{*\phi}_0,g_0)$.    

It will be useful to give explicitly some of the tensor decompositions
in the $G_2$ case. For instance, for two-forms one finds
\begin{equation}\label{twoforms}
   \Lambda^2= \Lambda^2_7\oplus\Lambda^2_{14}
\end{equation}
where 
\begin{equation}\label{reps}
\begin{aligned}
   \Lambda^2_7 &= \{\a \in \Lambda^2: *(\phi\wedge\a)=-2\a\}
       = \{i_{\b} \phi: \beta \in TM\} , \\
    \Lambda^2_{14}&=\{\a \in \Lambda^2: *(\phi\wedge\a)=\a\} .
\end{aligned}
\end{equation}
Recall that the space of two-forms $\Lambda^2$ is isomorphic
to the Lie algebra or adjoint representation $so(7)$. Thus this
decomposition is just the decomposition of $so(7)$ under $G_2$, namely
${\bf 21}\to {\bf 14}+{\bf 7}$, where ${\bf 14}$ is the Lie algebra
$g_2\cong\Lambda^2_{14}$ of $G_2$. There is, similarly, a
decomposition of three-forms 
\begin{equation}\label{threeforms}
   \Lambda^3= \Lambda^3_1\oplus\Lambda^3_7\oplus\Lambda^3_{27}
\end{equation}
following  $\mathbf{35}\to\mathbf{1}+\mathbf{7}+\mathbf{27}$ under
$G_2\subset SO(7)$. Note that elements of the singlet $\Lambda^3_1$
module are simply multiples of the $G_2$-invariant three-form $\phi$. 

Riemannian manifolds with $G_2$-structures have been classified
some time ago by Fernandez and Gray~\cite{Gray}. The idea is the
same for any $G$-structure on a Riemannian manifold, as discussed for
example in~\cite{sal}. Given some $G$-invariant form $\eta$ defining
a $G$-structure, the derivative of $\eta$ with respect to the
Levi--Civita connection,  $\nabla\eta$, can be decomposed into
$G$-modules. The different types of $G$-structures are then specified
by which of these modules are present, if any. 

In more detail, one first uses the fact that there is no obstruction to 
finding some connection $\nabla'$ such that $\nabla'\eta=0$. 
Choosing one, then $\nabla-\nabla'$ is a tensor which has values
in $\Lambda^1\otimes \Lambda^2$. Since $\Lambda^2\cong so(n)=g\oplus g^\perp$
where $g^\perp$ is the orthogonal complement of the Lie algebra $g$ in $so(n)$,
we conclude that $\nabla\eta =(\nabla-\nabla')\eta$ can be identified with
an element $K$ of $\Lambda^1\otimes g^\perp$. Futhermore, this element
is a function only of the particular $G$-structure, independent of the
choice of $\nabla'$. It is in one-to-one correspondence with what is
known as the intrinsic torsion $T_0$. Explicitly, we have in
components, acting on a $q$-form 
%
%\begin{equation}\label{Tdef}
%   \nabla \eta = - T \cdot \eta
%\end{equation}
%
%or in components $\nabla_m\eta_{n_1\dots n_q}=
%-T_{m\,n_1}{}^p\eta_{pn_2\dots n_q}-T_{m\,n_2}{}^p\eta_{n_1p\dots n_q}
%-\dots -T_{m\,n_q}{}^p\eta_{n_1\dots n_{q-1}p}$,
\begin{equation}\label{Tdef}
\nabla_m\eta_{n_1\dots n_q}=
-K_{m\,n_1}{}^p\eta_{pn_2\dots n_q}-K_{m\,n_2}{}^p\eta_{n_1p\dots n_q}
-\dots -K_{m\,n_q}{}^p\eta_{n_1\dots n_{q-1}p},
\end{equation}
where for $K_{m\,n}{}^p\in\Lambda^1\otimes g^\perp$, the $m$
and antisymmetric $n,p$ indices label the one-form $\Lambda^1$ and
two-form $g^\perp\subset\Lambda^2$ modules respectively. 

In the $G_2$ case, from the decomposition~\eqref{twoforms}, we see
that $g_2^\perp\cong\Lambda^2_7$, while $\Lambda^1$ is simply the
$\rep{7}$ representation of $G_2$. Thus specifying $\nabla\phi$ is
equivalent to giving elements in the four $G_2$-modules in the
decomposition of $K$ 
\be\label{decomp} 
   {\bf 7}\times {\bf 7} \to {\bf 1} + {\bf 7} + {\bf 14} +{\bf 27} 
\ee 
Given the general relation~\eqref{Tdef} with $\eta=\phi$, we see that
$d\phi$ and $d^{\dag}\phi\equiv -*d*\phi$ pick out different parts in
this decomposition. For example, following~\eqref{twoforms}, the
two-form $d^{\dag}\phi$ contains the ${\bf 7} + {\bf 14}$ pieces as 
follows 
\be\label{daggerphi}
   d^{\dag}\phi = i_\theta\phi + \a_{14} .
\ee 
Here $\a_{14}\in \Lambda^2_{14}$ and $\theta$ corresponds to the
$\rep{7}$, is called the Lee form and is given by
\be\label{lee}
   3\theta \equiv *(d^{\dag}\phi\wedge {*\phi}) ,
\ee 
or in components $\theta_a=-\frac{1}{6}\phi_{abc}\nabla_e\phi^{ebc}$.
Similarly, the four-form $d\phi$ can be decomposed into ${\bf 1}
+ {\bf 7} +{\bf 27}$ pieces, and so contains all but the ${\bf 14}$
in~\eqref{decomp}. Note, in particular, since it is derived from
the same general expression~\eqref{Tdef}, the ${\bf 7}$ in this
decomposition must be proportional to the Lee form defined in the
decomposition of $d^{\dag}\phi$.

It is clear from the above discussion that 
$\nabla^0\equiv\nabla + K$ canonically defines a connection for which 
$\nabla^0\eta =0$. It is the unique connection with torsion given by
the intrinsic torsion $T_0$. Since the holonomy of this connection,
and any connection $\nabla'$ for which $\nabla'\eta=0$, must
stabilise $\eta$ we conclude that its holonomy, $\Hol(\nabla^0)$,
must be contained within $G$. On the other hand demanding that
\textit{specific types} of connection have holonomy in
$G$, in general restricts the $G$-structure. For example, for the
Levi--Civita connection to have $\Hol(\nabla)\subseteq G$ we require
that all the elements in the decomposition of $K$ vanish so that
$\nabla\eta=0$. The $G$-structure is then said to be ``torsion-free''.    

This is probably the most familiar case of $G_2$-structure. With
torsion-free structure, so $\nabla\phi=0$, 
$(M_7,\phi,g)$ is said to be a ``$G_2$ manifold''.  It means that the
Levi--Civita connection $\nabla$ has holonomy contained in $G_2$ and
$g$ is a Ricci-flat metric. Given the preceding discussion it is clear
that the condition $\nabla\phi=0$ is equivalent to requiring
\be\label{g2manifolds}
   d\phi=d{*\phi}=0
\ee 
since all the relevant $G_2$-modules in $\nabla\phi=0$ appear either
in $d\phi$ or $d{*\phi}$. This equivalence has been exploited
in~\cite{Brand} to provide a method for finding $G_2$ holonomy metrics
for manifolds of co-homogeneity one. The strategy is the
following. Write down an ansatz for the associative three form $\phi$
in terms of several arbitrary functions of one radial variable. Find
the associated metric and impose the conditions~\p{g2manifolds} 
to obtain a system of first-order differential equations for the
arbitrary functions. Solving these one obtains a $G_2$ holonomy
metric.

For type II supergravity solutions describing NS fivebranes wrapping 
supersymmetric three-cycles one finds seven manifolds
with $G_2$ structures of a different type~\cite{GKMW} since
the connection with holonomy in $G_2$ is not the Levi--Civita
connection $\nabla$. This is reviewed in the next section. We can
derive an analogous pair of necessary and sufficient conditions
to~\eqref{g2manifolds}. We will then exploit these,
generalising~\cite{Brand}, to find new solutions in a later section.

\section{$G_2$-structure and NS fivebranes on associative three-cycles}

The action for the bosonic NS-NS fields of type IIA or type IIB
supergravity is given by 
\be 
   S = \frac{1}{2\kappa^2}\int d^{10}x \sqrt {-g} e^{-2\Phi}
     \left(R+4(\nabla\Phi)^2-\frac{1}{12}H^2\right) ,
\ee 
with $H=dB$. The corresponding equations of motion read
\begin{equation}
\label{eom}
\begin{aligned}
   R_{\mu\nu} - \frac{1}{4}H_{\mu\rho\sigma}H_{\nu}{}^{\rho\sigma}
      + 2\nabla_\mu\nabla_\nu\Phi &= 0 , \\
   \nabla^2\Phi - 2(\nabla\Phi)^2 
      +\frac{1}{12}H_{\mu\nu\rho}H^{\mu\nu\rho} &= 0 , \\
   \nabla^\mu \left( e^{-2\Phi} H_{\mu\nu\rho} \right) &= 0 .
\end{aligned}
\end{equation}

As shown in~\cite{agk}, the IIB supergravity solution
describing fivebranes wrapped on an associative three-cycle in a
manifold of $G_2$-holonomy is of the form $\RR^{1,2}\times M_7$
where $M_7$ admits a single $Spin(7)$ spinor satisfying
\begin{equation}
\label{G2spinor}
\begin{gathered}
    \nabla^{+}_m\e
        \equiv \left(\nabla_m+\frac{1}{8}H_{mnp}\gamma^{np}\right)\e
        = 0 , \\
    H_{mnr}\gamma^{mnr}\e = -12\partial_{n}\Phi\gamma^n\e ,
\end{gathered}
\end{equation}
where  $\nabla^{+}$ is a connection with totally antisymmetric
torsion $\frac{1}{2}H$. Here the $Spin(7)$ Dirac matrices $\gamma^m$ are
imaginary, antisymmetric and satisfy 
\be 
\{\gamma_m,\gamma_n\}=2\delta_{mn} 
\ee 
with $\gamma^1\cdots\gamma^7=-i$.

From the existence of a single solution to the first equation in
\p{G2spinor} it immediately follows that we have a $G_2$-structure on $M_7$
and the $\Hol(\nabla^{+})=G_2$. To see this, we first 
define a three form $\phi$ in terms
of the Killing spinor by:
\bea\label{phi}
\phi_{mnr}=i\e^T\gamma_{mnr}\e
\eea
where we have normalized the spinors to satisfy $\e^T\e=1$. This
$G_2$-structure then satisfies
\be
\nabla^+\phi=0
\ee
and hence $\Hol(\nabla^{+})\subseteq G_2$. Using 
the second equation in \p{G2spinor} we find that the $G_2$-structure
also satisfies
\bea
\label{relations}
   -*e^{2\Phi}d(e^{-2\Phi}\phi)&=&H \nn
   d(e^{-2\Phi}{*\phi})&=&0\nn
d\phi\wedge\phi&=&0
\eea
That is, for a solution to the equations of motion 
to preserve supersymmetry, $M_7$ must admit a 
$G_2$ structure which satisfies the conditions \p{relations} with $H$ closed.
This form of the conditions \cite{GKMW} naturally displays the connections
with generalised calibrations \cite{GPT}.

A converse result has also been proved in \cite{Ivanov1,Ivanov2}.
Let us summarise the idea behind it before extending it. One assumes the
existence of a $G_2$-structure satisfying the last two equations
in \p{relations}. Recalling the definition of the Lee-form introduced in
\p{daggerphi} it is easy to see
that $d(e^{-2\Phi}* \phi)=0$ is equivalent to the
statement that (i) the Lee form is given by $\theta=-2d\Phi$
and (ii) that $\a_{14}$ vanishes. It was shown in \cite{Ivanov1}
that the second condition is the necessary
and sufficient condition for the existence of a unique connection
$\nabla^{+}=\nabla+\frac{1}{2}H$, with totally anti-symmetric torsion 
$\frac{1}{2}H$, that preserves the three form
$\phi$ and admits parallel spinors. The idea behind this is rather simple.
First recall
from~\eqref{threeforms} that $H$, which is in the $\rep{35}$ of
$SO(7)$, decomposes under $G_2$ as ${\bf 35}\to {\bf 1}+{\bf 7}
+{\bf 27}$. On the other hand as we discussed in section 2, the
different types of $G_2$ structure correspond to the modules given in
the decomposition~\p{decomp} of $K$. It is thus clear that totally
anti-symmetric torsion is associated with vanishing
$\a_{14}$ in $K$. Moreover, it was shown in \cite{Ivanov2} that the $G_2$
singlet piece in $H$ is proportional to
$*(d\phi\wedge\phi)$ and vanishes if and only if the supersymmetry
variation of the dilatino vanishes. The point is that the Clifford
action of the ${\bf 27}$ piece of $H$ on $\e$ vanishes. 

In other words it was proved in \cite{Ivanov2} that a manifold with a
$G_2$ structure $(M_7,g,\phi)$ admits solutions to \p{G2spinor} with
varying dilaton and non vanishing NS three form $H$ providing that the
$G_2$ structure satisfies:
\bea\label{conditions}
 d\phi\wedge\phi&=&0\nn
~ *(\phi\wedge d^\dagger\phi)&=&-2d^\dagger\phi\nn 
 \theta&=&-2d\Phi
\eea
or equivalently
\bea\label{conditionsalt}
d\phi\wedge \phi&=&0\nn
d(e^{-2\Phi}* \phi)&=&0
\eea
The torsion of the unique connection with totally anti-symmetric 
torsion preserving the $G_2$ structure is then given by 
$H=-*e^{2\Phi}d(e^{-2\Phi}\phi)$.

Note that supersymmetry alone is not sufficient to ensure that we
have a solution to the type II field equations. We also need to
impose at least the closure of $H$. In fact this is all we need as we
now show using the integrability conditions of the Killing spinor
equations~\eqref{G2spinor}. As shown in the Appendix these imply that 
\begin{equation}\label{othereq} 
\begin{aligned}
   \bigg( R_{mn} - \frac{1}{4}H_{mpq}H_{n}{}^{pq} 
       &+ 2\nabla_m\nabla_n\Phi \bigg)\gamma^n\epsilon \\
   &= \frac{1}{12} dH_{mnpq} \gamma^{npq}\epsilon
       + \frac{1}{2}e^{2\Phi}
           \nabla^p\left(e^{-2\Phi}H_{pmn}\right)\gamma^n\epsilon
\end{aligned}
\end{equation}
and
\begin{equation}\label{phieq}
\begin{aligned}
   \bigg( \nabla^2\Phi - 2[\nabla\Phi]^2
       &+ \frac{1}{12}H_{mnp}H^{mnp} \bigg) \epsilon \\
   & = - \frac{1}{48} dH_{mnpq}\gamma^{mnpq} \epsilon
       - \frac{1}{4}e^{2\Phi}
           \nabla^m\left(e^{-2\Phi}H_{mnp}\right)\gamma^{np}\epsilon
\end{aligned}
\end{equation}
The assumptions on the $G_2$ structure~\eqref{conditionsalt}
mean that the $H$ equation  of motion~\eqref{eom} is automatically
satisfied. We thus immediately conclude from~\eqref{phieq} that, if we
also impose $dH=0$, then the dilaton equation of motion is
satisfied. The other equation~\eqref{othereq} is then of the form
$A_{mn}\gamma^n\e=0$ which  implies  $A_{mn}A^{mn}=0$. On a Riemannian
manifold we then deduce $A_{mn}=0$ which is precisely the Einstein
equations. 

In summary, we have shown that a solution of the equations of motion
of the form $\RR^{1,2}\times M_7$ admits a single Killing spinor if and only
if
\begin{equation}
\label{conditions2} 
\begin{aligned} 
   d\phi\wedge \phi &= 0 , \\
   d(e^{-2\Phi}*\phi) &= 0 , \\
   dH &= 0 ,
\end{aligned}
\end{equation}
where $H=-*e^{2\Phi}d(e^{-2\Phi}\phi)$. This result is the analog
of~\eqref{g2manifolds} for $G_2$-manifolds and in principle provides a
method for finding new supersymmetric solutions with non-zero $H$. One
starts with an ansatz for $\phi$, finds the associated metric and
imposes these equations to obtain, in the case of a metric of
co-homogeneity one, ordinary differential equations for the arbitrary
metric functions. We give examples of this technique in
section~\ref{sec:sol1}.  

It is interesting to note that the expression for $H$ implies a simple
vanishing 
theorem\footnote{A different vanishing theorem was proved in \cite{agricola}, 
which assumed vanishing dilaton.}: on a compact manifold without boundary the only solutions 
to~\eqref{conditions2} have $H=0$ and $\Phi$ constant, that is, $M_7$
is a $G_2$-manifold. To see this, we first note, given
the expression for $H$ 
\begin{equation}
   \int_{M_7}e^{-2\Phi}H\wedge *H 
       = -\int_{M_7} H\wedge d(e^{-2\Phi}\phi) = 0 ,
\end{equation}
where in the final equation we integrate by parts and use
$dH=0$. Since the first integrand is positive definite, we conclude
that $H=0$. Integrating the dilaton equation of motion
then implies by a similar argument
that $d\Phi=0$ so $\Phi$ is constant. The
conditions~\eqref{conditions2} then reduce to $d\phi=d{*\phi}=0$ which
imply that $M_7$ is a $G_2$-manifold. In~\cite{GKMW} we derived
analogous expressions for $H$ in terms of the generalised calibrations
for other geometries in dimensions six and eight
arising when fivebranes wrap calibrated
cycles. Since only this expression and the $\Phi$ equation of motion
entered the above argument, clearly this theorem easily
generalises. For all compact supersymmetric manifolds $M$
without boundary, the flux
vanishes $H=0$ and $\Phi$ is constant.

\section{$SU(3)$-structure and NS fivebranes on SLAG three-cycles}

It was shown in~\cite{GKMW} that the type II supergravity
solutions describing fivebranes wrapping SLAG three-cycles are also of
the form $\RR^{1,2}\times M_7$ where now $M_7$ admits a pair of
$Spin(7)$ spinors $\epsilon^\pm$ satisfying 
\begin{equation}
\label{su3spinoreq}
\begin{gathered}
   \nabla^{\pm}_m\e^{\pm} \equiv 
      \left(\nabla_m\pm\frac{1}{8}H_{mnp}\gamma^{np}\right)\e^{\pm}
      = 0 , \\
   H_{mnr}\gamma^{mnr}\e^{\pm} = \mp12\partial_{n}\Phi\gamma^n\e^{\pm} ,
\end{gathered}
\end{equation}
where $\nabla^{\pm}$ are two connections with anti-symmetric
torsion $\pm\frac{1}{2}H$. From the discussion of the previous
section, it follows that the two spinors $\epsilon^\pm$ define two
distinct $G_2$-structures of the type characterised by the
conditions~\eqref{conditions2}. The two connections $\nabla^\pm$ have
holonomy contained in two different $G_2$ subgroups of $SO(7)$. 
The $G_2$-invariant three forms are constructed from the Killing
spinors
\be 
   \phi^{\pm}_{mnr}=i\e^{\pm T}\gamma_{mnr}\e^{\pm} ,
\ee 
where we have again chosen the normalization $\e^{\pm T}\e^{\pm}=1$,
as we can always do. 

The appearance of two $G_2$-structures is again quite general,
depending only on the requirement that there are two distinct
solutions $\epsilon^+$ and $\epsilon^-$. We can analyse this structure
further, making only one further assumption, as in~\cite{GKMW}, that
the the Killing spinors are orthogonal to each other \textit{i.e.}
$\e^{+T}\e^{-}=0$. We expect that this should cover the general class
of supersymmetric solutions describing fivebranes wrapped on SLAG
three-cycles. This is because we can deduce what projections we expect
to be imposed on the preserved supersymmetries by considering the
supersymmetries preserved by a fivebrane probe wrapped on a SLAG
three-cycle, as described in \cite{GKMW}. Note in particular, that
this condition was satisfied for the specific supersymmetric
supergravity solutions of \cite{gr,GKMW}. It is equivalent to the
statement that the two $G_2$ structures satisfy
\be\label{ortho} 
   \phi^{+}_{[m}{}^{r_1r_2}\phi^{-}{_{n]r_1r_2}}=0 ,
\ee
as can be shown by Fierz rearrangement.

Apart from the two $G_2$ three-forms we can also construct various
other forms using the Killing spinors
\begin{equation}
\begin{aligned}
   K_n &= i\e^{+T}\gamma_n\e^{-} , \\
   \omega_{mn} &= \e^{+T}\gamma_{mn}\e^{-} , \\
   \chi_{mnr} &= i\e^{+T}\gamma_{mnr}\e^{-} . 
\end{aligned}
\end{equation}
These are the basic objects in the sense that the two $G_2$ structures
can be constructed from $K$, $\omega$ and $\chi$ as follows
\be\label{g2fromsu3} 
    \phi^{\pm}=\pm i_{K}*\chi- K\wedge\omega . 
\ee

$K,\omega,\chi$ satisfy a series of algebraic relations that
follow from Fierz rearrangements. First, given the normalization
of the Killing spinors, we find that: 
\begin{equation}
\begin{aligned}
   K_mK^m &=1 , \\
   \omega_{mn}\omega^{mn} &= 6 , \\
   \chi_{mnp}\chi^{mnp} &=24 ,
\end{aligned}
\end{equation}
and also
\begin{equation}
\label{flip} 
\begin{aligned}
   \omega_{m}{^r}\omega_r{^n} &= -\delta^n_m+K_m K^n , \\
   i_K \chi &= 0 ,\\\
   i_K \omega &= 0 , \\
   (i_K*\chi)_{mnr} &= \omega_{m}{^l}\chi_{nrl} 
       = \omega_{[m}{^l}\chi_{nr]l} .
\end{aligned}
\end{equation}

In addition, we can calculate the covariant derivatives of these forms
using the first Killing spinor equation to get
\begin{equation}
\begin{aligned}
   \nabla_m K_n &= \frac{1}{4}H_{m l_1 l_2}\chi_{n}{^{l_1l_2}}, \\
   \nabla_m \omega_{n_1n_2} &= 
       \frac{1}{4}H_{ml_1l_2}{*\chi}_{n_1n_2}{^{l_1l_2}} , \\
   \nabla_m \chi_{n_1n_2n_3} &= -\frac{3}{2}H_{m[n_1n_2}K_{n_3]}
       -\frac{1}{4}H_{ml_1l_2}{*\omega}_{n_1n_2n_3}{^{l_1l_2}} .
\end{aligned}
\end{equation}
From the dilatino equations we then deduce the following relations for
the exterior derivatives of the forms
\begin{equation}
\begin{aligned}
   d(e^{-\Phi}K) &= 0 , \\
   d(e^{-\Phi}\omega) &= 0 , \\
   e^{\Phi}d(e^{-\Phi}\chi) &= -H \wedge K , 
\end{aligned}
\end{equation}
as well as for the $G_2$-structures as in~\eqref{relations},
\begin{equation}
      e^{2\Phi}d(e^{-2\Phi}\phi^{\pm}) = \mp *H .
\end{equation}
It is also not difficult to show in addition that 
\begin{equation}
\begin{aligned}
   d(i_{K}*\chi)\wedge i_{K}*\chi &= 0 , \\
   d(i_{K}*\chi)\wedge K\wedge \omega &= 0 .  
\end{aligned}
\end{equation}

\subsection{$SU(3)$ structure}

Let us now discuss what the presence of these three invariant forms
$K$, $\omega$ and $\chi$ implies about the type of a $G$-structure we
have on $M_7$. Recall that the existence of $\phi^\pm$, or
equivalently $\epsilon^\pm$, implied that there were two distinct 
$G_2$-structures. Of course there can only be one actual
structure group $G$ of the frame bundle, so the implication is that
$G$ must be a common subgroup of these two distinct embeddings of
$G_2$ in $SO(7)$. The largest possible such group is $SU(3)$. This is
consistent with the existence of two Killing spinors since the
$\rep{8}$ spinor of $Spin(7)$ includes two singlets in its
decomposition  $\rep{1}+\rep{1}+\rep{3}+\rep{\bar{3}}$ under
$SU(3)$. Thus we might expect that in fact there is actually an
$SU(3)$-structure on $M_7$. By considering each of the invariant
forms $K$, $\omega$ and $\chi$ in turn we will see that this is indeed
the case. Each will further restrict the $G$-structure until we are
left with $SU(3)$.

We start with $K$. Clearly, a fixed vector is left invariant by
$SO(6)\subset SO(7)$ rotations of the orthonormal frame. Thus we
see that $K$ defines an $SO(6)$-structure. Equivalently we can
introduce  
\be
    \Pi_{m}{^n} = 2K_m K^n-\delta_{m}{^n} 
\ee
satisfying $\Pi^2=\delta$, since $K^2=1$, and hence defining an almost
product structure. It is metric compatible in the sense that $\Pi
g\Pi^T=g$, or equivalently $\Pi_{mn}=\Pi_m{^r}g_{rn}$ is symmetric. It
is also integrable in that its Nijenhuis tensor defined by  
\be 
   N_{mn}{^r} = \Pi_{m}{^k}\partial_{[n}\Pi_{k]}{^r}
      - \Pi_{n}{^k}\partial_{[m}\Pi_{k]}{^r}  
\ee
vanishes using $d(e^{-\Phi}K)=0$. This implies that we in fact have a
product structure. It follows that we can find coordinates such that
$\Pi$ is diagonal, or equivalently $K=e^\Phi dx^7$. In these
coordinates the seven-dimensional metric takes the form  
\be\label{metricdecomp}
   ds_7^2=g_{ab}dx^adx^b+e^{2\Phi}dx_7^2 .
\ee
Note that the geometry is not a direct product since $g_{ab}$ and
$\Phi$ are allowed to depend on all the coordinates. The metrics of
the solutions presented in~\cite{gr,GKMW} indeed have this form, as we shall
show in section 6.  

Now consider $\omega$. The pair $(K,\omega)$ define what is known as
an almost contact metric structure (see for
example~\cite{blair}). This means, in general, that the structure
group of the frame bundle on a manifold $M_{2k+1}$ reduces from
$SO(2k+1)$ to $U(k)$, implying here that we have an
$U(3)$-structure. It is the analog of an almost hermitian structure
for odd-dimensional manifolds. A manifold $M_{2k+1}$ is said to have
an almost contact 
metric structure if it admits a $(1,1)$ tensor $\omega_m{}^n$ and a
one-form $K$ satisfying the first equation of~\p{flip}, and
furthermore $\omega$ is metric compatible so that $\omega
g\omega^T=g$, or equivalently $\omega_{mn}=\omega_m{^r}g_{rn}$ is a
two form. Note that this implies $i_K\omega=0$. Essentially, the
existence of $K$ allows one to consistently decompose the tangent
space into $2k$-dimensional piece and a one-dimensional piece. The
two-form $\omega$ then defines an almost hermitian structure on the
$2k$-dimensional piece, so that the corresponding complexified tangent
space splits into the sum of a $k$-dimensional complex space and its
complex conjugate. Thus, in general, we have the decomposition
$TM_{2k+1}\otimes\CC=T^{1,0}\oplus T^{0,1}\oplus
(T_K\otimes\CC)$. There is an integrability condition similar to that
of an almost hermitian structure and if it is integrable the almost
contact metric structure is called normal. In the geometries we are
interested in the structure is not integrable in general.  

It is interesting to note that restricting to the six-dimensional part
of the metric~\p{metricdecomp} by setting $x_7$ constant we obtain a
conventional almost hermitian structure. However this is again not
integrable and the six-dimensional manifold is not a complex
manifold. This is perhaps surprising since such solutions describe
fivebranes wrapped on SLAG three cycles in Calabi-Yau manifolds and hence
one might naively have expected that the six-dimensional part of
the geometry is complex.

Finally we come to $\chi$. We first note that we can define a complex
three-form
\be 
   \vartheta = \chi - i(i_K {*\chi}) .
\ee
We see that $\vartheta$ is normal to $K$, and also, from the third
equation of~\p{flip}, that it is a $(3,0)$-form with respect to the
almost contact structure $(K,\omega)$. In other words it is a section
of $\Lambda^3T^{1,0}$. In this sense it is the analog of the holomorphic
three-form on the original Calabi--Yau manifold. In fact, in an
exactly analogous way, it is easy to see that the subgroup of $U(3)$ 
which preserves $\vartheta$ is $SU(3)$. Thus we conclude that we do
indeed have an $SU(3)$-structure on $M_7$. 

Since $d(e^{-\Phi}K)=d(e^{-\Phi}\omega)=0$ it follows that
\bea\label{su3H}
   * H = e^{2\Phi}d(e^{-2\Phi}\im\vartheta)
\eea 
which shows that $\vartheta$ is a generalized calibration.
This mirrors the fact that for a Calabi-Yau 3-fold the imaginary
part of the holomorphic three-form calibrates SLAG three-cycles.

\subsection{The necessary and sufficient conditions}

We have shown that given two spinors $\epsilon^\pm$
satisfying~\p{su3spinoreq}, with $\epsilon^{+T}\epsilon^-=0$, $M_7$
necessarily has an $SU(3)$-structure given by $(K,\omega,\chi)$. The
structure is not general, but as in the case of
$G_2$-structure above is restricted. We have seen already that, for
instance, the almost product structure defined by $K$ is integrable,
though the almost contact structure $(K,\omega)$ is not. In general,
we showed that we have the conditions 
\begin{equation}\label{su3conditions}
\begin{gathered}
   dK=d\Phi\wedge K \\
   d\omega=d\Phi\wedge \omega \\
   d\chi\wedge K=d\Phi\wedge\chi\wedge K \\
   d(i_{K}*\chi)\wedge i_{K}*\chi=0 \\
   d(i_{K}*\chi)\wedge K\wedge \omega=0 
\end{gathered}
\end{equation}
These are also sufficient conditions for the existence of a solution
to~\eqref{su3spinoreq}. To see this one needs to show
that $\phi^\pm$ defined by~\p{g2fromsu3} satisfy~\p{ortho}, which
follow from the algebraic properties of $(K,\omega,\chi)$ and also
the each satisfy the conditions~\p{conditions2}. The latter is
straightforward to show using the fact that ${*\phi}^{\pm}=\pm
\chi\wedge K-\frac{1}{2}\omega\wedge\omega$.

In other words, a spacetime of the form $\RR^{1,2}\times M_7$ admits
two orthogonal Killing spinors  $\epsilon^\pm$
satisfying~\eqref{su3spinoreq}, if and olnly if
we have an $SU(3)$-structure on
$M_7$ satisfying the above conditions~\eqref{su3conditions}. In
addition, the antisymmetric torsion $\pm\frac{1}{2}H$ is given
by~\p{su3H}. Furthermore, this gives a supersymmetric solution of type
II supergravity if and only if we impose in addition the closure of $H$ as
before.  

Once again we can in principle use this result as a method for
finding solutions. In section 6 we shall recover the solutions
that were found in \cite{gr,GKMW} using gauged supergravity
techniques.

\section{Constructing solutions with a single $G_2$ structure}
\label{sec:sol1}

In this section we will use the results of section 3 to construct
examples of the geometries with a single $G_2$-structure that
were described there. These correspond to fivebranes wrapping
associative three-cycles. We focus on co-homogeneity one
manifolds.

We generalise the method presented in \cite{Brand}.
One first makes an ansatz for the $G_2$ structure $\phi$
satisfying appropriate symmetries and then finds the associated
metric from the expression: 
\bea\label{assocmetric}
g_{ij}&=&(det~s_{ij})^{-1/9}s_{ij} \nn
s_{ij}&=&\frac{1}{144}\phi_{i n_1n_2}\phi_{j n_3n_3}\phi_{n_5 n_6
n_7}\e^{n_1n_2n_3n_4n_5n_6n_7}, ~~ \e^{1234567}=1 
\eea 
The three form $\phi$ must be {\it stable} in the sense of
\cite{Hitchin} to ensure that it is generic enough to make the metric
non-degenerate. We then impose equations \p{conditionsalt}. If these
are satisfied we have a solution to \p{G2spinor}. One must then
impose the closure of $H$ to obtain a solution to the full
supergravity theory.

\subsection{The example of \cite{agk}}
Let us first demonstrate this method by recovering the example
presented in \cite{agk}. This is a co-homogeneity one example
with principle orbits $S^3\times S^3$. Our starting point is an
ansatz for the three form that has appeared in constructions of
$G_2$ holonomy metrics in \cite{Gibb1},\cite{Gukov}:
\begin{multline}\label{ansatz}
\phi=a b~ dt\wedge\sum_{a=1}^3(\Sigma_a-\frac{1}{2}\sigma_a)\wedge
\sigma_a+\frac{a^3}{3!}\e_{abc}~\sigma_a\wedge\sigma_b\wedge\sigma_c\\
-\frac{ab^2}{2!}\e_{abc}\sigma_a\wedge(\Sigma_b-\frac{1}{2}\sigma_b)\wedge(\Sigma_c-\frac{1}{2}\sigma_c)
\end{multline}
where $(\Sigma_a,\sigma_a)$ are left invariant one-forms on
$SU(2)\times SU(2)$, satisfying
$d\sigma_1=-\sigma_2\wedge\sigma_3$,
$d\Sigma_1=-\Sigma_2\wedge\Sigma_3$ plus cyclic permutations, and
$t$ is a radial variable. The two arbitrary functions $a$ and $b$
depend on the radial variable only. The associated metric is given
by
\bea\label{metric}
ds_7^2=dt^2+b^2\sum_{a=1}^3(\Sigma_a-\frac{1}{2}\sigma_a)^2+a^2\sum_{a=1}^3(\sigma_a)^2
\eea

Introduce a frame
 \bea\label{frame} 
e^t&=&dt\nn
e^a&=&b(\Sigma_a-\frac{1}{2}\sigma_a)\nn \tilde{e}^a&=&a~\sigma_a
\eea 
In this frame the three form $\phi$ and its dual are given by (with
$\e^{t123\tilde{1}\tilde{2}\tilde{3}}=-1$)
\begin{equation}\label{framephi}
\begin{aligned}
   \phi &= e^t\wedge e^a\wedge\tilde{e}^a 
       + \frac{1}{3!}\e_{abc}\,\tilde{e}^a\wedge\tilde{e}^b\wedge\tilde{e}^c
       - \frac{1}{2!}\e_{abc}\,\tilde{e}^a\wedge e^b\wedge e^c \\
   *\phi &= -\frac{1}{3!}\e_{abc}\,e^t\wedge e^a\wedge e^b\wedge e^c
       + \frac{1}{2!}\e_{abc}\,e^t\wedge
          e^a\wedge\tilde{e}^b\wedge\tilde{e}^c
       + \frac{1}{2!}\tilde{e}^a\wedge e^a\wedge\tilde{e}^b\wedge e^b
\end{aligned}
\end{equation}
and it is straightforward to calculate: 
\bea
d\phi&=&\frac{1}{2}\{(\log a)'-\frac{b}{4a^2}\}\e_{abc}~e^t\wedge
\tilde{e}^a\wedge\tilde{e}^b\wedge\tilde{e}^c-\frac{1}{2}\{(\log
b^2a)'-\frac{1}{b}\}\e_{abc}~e^t\wedge \tilde{e}^a\wedge e^b\wedge
e^c \nn d\ast\phi&=&\{(\log
ab)'-\frac{1}{2b}-\frac{b}{8a^2}\}~e^t\wedge \tilde{e}^a\wedge e^a
\wedge \tilde{e}^b \wedge e^b 
\eea

First note that $d\phi\wedge \phi=0$  is automatically satisfied.
Also we have $d\ast\phi=d(2\Phi)\wedge \ast\phi$ with:
\be\label{dilaton} 
    d(2\Phi)=\left[(\log a^2b^2)'
        - \frac{1}{b} - \frac{b}{4a^2} \right] dt 
\ee 
So all the conditions
\p{conditions} are satisfied and we have a solution to the
Killing spinor equations \p{G2spinor}. Note that the two
functions $a,b$ are still arbitrary. This is because we started
with a very special ansatz which guaranteed from the beginning
that all the conditions were satisfied. However we still need to
impose the closure of H. This will give us second order equations
in principle but as we shall see, in this case they are trivially
integrated once. The torsion $H$ is constructed from \p{relations}
and we find that: 
\bea\label{torsion} H=\frac{1}{3!}\e_{abc}F~
e^a \wedge e^b\wedge e^c+\frac{1}{2!}\e_{abc}G~e^a\wedge
\tilde{e}^b \wedge \tilde{e}^c 
\eea 
where 
\bea F&=&\{(\log b^2
a^{-1})'-\frac{1}{b}+\frac{b}{2a^2}\}\nn G&=&\{-(\log
a)'+\frac{b}{4a^2}\} 
\eea 
Imposing $dH=0$ we get the equations
\bea\label{bpseqn}
F'+F(\log b^3)'&=&0\\
G'+G(\log ba^2)'&=&0\\
G\frac{1}{2b}-F\frac{b}{8a^2}&=&0 
\eea 
The first two are
trivially integrated to give $F b^3=C_1$ and $G ba^2=C_2$ while
the third implies that $C_1=4C_2\equiv -\mu$. Using the
definitions of $F$ and $G$ we thus arrive at a system of first
order equations for the metric functions $a,b$:
\bea\label{diffeqns}
b'&=&\frac{1}{2}(1-\frac{\mu}{b^2})(1-\frac{b^2}{4a^2})\\
a'&=&\frac{b}{4a}(1+\frac{\mu}{b^2}) \eea

These equations are precisely those derived at the end of section
3.1.1 of \cite{agk}. For the special case $\mu=0$ the torsion and
the dilaton vanish and the equations can be integrated to
recover the $G_2$ holonomy metric on the spin bundle of $S^3$
\cite{bs}. The solution with $b^2=\mu$, $a^2={\sqrt \mu} r$ was
found in \cite{agk} using gauged supergravity methods. It
corresponds to fivebranes wrapped on the associative three-sphere
of the $G_2$-holonomy manifolds of \cite{bs}, in the near horizon
limit. The general solution of these equations remains an
outstanding problem.

One can extend this analysis in a relatively straightforward way
to recover the solution first presented in \cite{maldnast}, but
the formulae are rather lengthy so we shall not present the
details here.

\subsection{New Solution}

\par Another example is to start with a cohomogeneity one manifold with
principal orbits $SU(3)/U(1)\times U(1)$. Such $G_2$ structures
have appeared in \cite{Cleyton},~\cite{Gibb2}, and solutions have
been found for a $G_2$ metric on the $\mathbb{\RR}^3$ bundle over
$\mathbb{CP}^2$ \cite{bs}. Here we use the results of
\cite{Cleyton} to find and solve the BPS equations for solutions
that describe fivebranes wrapped on the $\mathbb{\RR}^3$ fibres, which are 
non-compact associative three cycles, in such $G_2$ manifolds.

\par Let $\{e_a\}$ be the left invariant one forms on $SU(3)$. We
define \bea\label{omegas} \omega_1=e_{12}, ~~~ \omega_2=e_{34},
~~~\omega_3=e_{56} \eea and also a basis for the $SU(3)$
invariant three forms: \bea\label{ab}
\a=e_{246}-e_{235}-e_{145}-e_{136},
~~~\b=e_{135}-e_{146}-e_{236}-e_{245} \eea where $e_{12}\equiv
e_1e_2$ etc, and the exterior product of forms is understood. It
then follows that these satisfy \cite{Cleyton}
\bea\label{extalgebra}
d\omega_1&=&d\omega_2=d\omega_3=\frac{1}{2}\a\nn d\a&=&0\nn
d\b&=&-2(\omega_1\wedge
\omega_2+\omega_2\wedge\omega_3+\omega_3\wedge\omega_1)\nn
d(\omega_i \wedge \omega_j)&=&0, ~~i\neq j \eea The $G_2$
structure and it's associated metric are given by:
\bea\label{phi2}
\phi=(f_1^2\omega_1+f_2^2\omega_2+f_3^2\omega_3)\wedge
dt+f_1f_2f_3(\cos\theta ~\a+\sin\theta ~\b) \eea
\bea\label{metric2} ds_7^2=dt^2+f_1^2g_1+f_2^2g_2+f_3^2g_3 \eea
where $f_i,~\theta$ are arbitrary functions of $t$ and
\bea\label{g's}
g_1=e_1^2+e_2^2,~~~g_2=e_3^2+e_4^2,~~~g_3=e_5^2+e_6^2 \eea We
find that \bea\label{starphi2} \ast
\phi=f_2^2f_3^2\omega_2\wedge\omega_3+f_3^2f_1^2\omega_3\wedge\omega_1+
f_1^2f_2^2\omega_1\wedge\omega_2+f_1f_2f_3(\cos \theta
~\b-\sin\theta~\a)\wedge dt \eea and hence \bea\label{dphi2}
d\phi&=&(\frac{1}{2}(f_1^2+f_2^2+f_3^2)-(f_1f_2f_3\cos\theta)')\a\wedge
dt\nn &-&(f_1f_2f_3\sin\theta)'\b\wedge dt\nn
&-&2f_1f_2f_3\sin\theta(\omega_1\wedge
\omega_2+\omega_2\wedge\omega_3+\omega_3\wedge\omega_1) \eea
\bea\label{dstarphi2}
d\ast\phi&=&((f_2^2f_3^2)'-2f_1f_2f_3\cos\theta)\omega_2\wedge\omega_3\wedge
dt\nn
&+&((f_3^2f_1^2)'-2f_1f_2f_3\cos\theta)\omega_3\wedge\omega_1\wedge
dt\nn&+&((f_1^2f_2^2)'-2f_1f_2f_3\cos\theta)\omega_1\wedge\omega_2\wedge
dt \eea

Let us first consider the equation $d\ast\phi=d(2\Phi)\wedge
\ast\phi$. This gives: \bea
\Phi'=\log(f_if_j)'-\frac{f_k}{f_if_j}\cos \theta, ~~~i\neq j\neq
k \eea which defines the dilaton and also imposes: \bea\label{eq1}
f_k \log f_i '+\cos\theta \frac{f_i}{f_j}=f_k \log f_j
'+\cos\theta \frac{f_j}{f_i},~~i\neq j\neq k \eea which is two
independent equations.
Next we impose $(\ast\phi,d\phi)=0$ and we conclude that:
\bea\label{eq2}
\theta'=-\sin\theta\frac{f_1^2+f_2^2+f_3^2}{f_1f_2f_3}
\eea
Having satisfied these conditions we need to impose the closure
of $H$. We find that the torsion is given by: 
\begin{equation}\label{torsion2}
\begin{split}
   H &= 2f_1f_2f_3\sin\theta \left(
                \frac{f_3^2}{f_1^2f_2^2}\omega_3
                + \frac{f_1^2}{f_2^2f_3^2}\omega_1
                + \frac{f_2^2}{f_1^2f_3^2}\omega_2 \right) \wedge dt 
      \\ & \qquad
        - \bigg[ 2\Phi'f_1f_2f_3\sin\theta 
           - (f_1f_2f_3\sin\theta)'\bigg]\a 
      \\ & \qquad
        - \left[ (f_1f_2f_3\cos\theta)'
           - 2\Phi'f_1f_2f_3\cos\theta -
                \frac{1}{2}(f_1^2+f_2^2+f_3^2) \right]\b  
\end{split}
\end{equation}
For $H$ to be closed we thus need to impose the following:
\bea\label{eq3}
(f_1f_2f_3\cos\theta)'-2\Phi'f_1f_2f_3\cos\theta-\frac{1}{2}(f_1^2+f_2^2+f_3^2)=0
\eea \bea\label{redeq}
[2\Phi'f_1f_2f_3\sin\theta-(f_1f_2f_3\sin\theta)']'+f_1f_2f_3\sin\theta(\frac{f_3^2}{f_1^2f_2^2}
+\frac{f_1^2}{f_2^2f_3^2}+\frac{f_2^2}{f_1^2f_3^2})=0 \eea

Now \p{redeq} is a second order equation and also we have five
equations for four unknown functions. However the second order
equation follows from the four first order equations so there is
no inconsistency. To see this we first rearrange equations
\p{eq1},\p{eq2},\p{eq3}, and write them as: \bea\label{fi'} (\log
f_i)'&=&\frac{1}{2\cos\theta}\frac{f_1^2+f_2^2+f_3^2}{f_1f_2f_3}-\cos\theta\frac{f_i^2}{f_1f_2f_3},~~i=1,2,3\nn
\theta'&=&-\sin\theta\frac{f_1^2+f_2^2+f_3^2}{f_1f_2f_3}
\eea Then we note that we can write \p{redeq} as : \bea
\frac{1}{2}(\tan\theta(f_1^2+f_2^2+f_3^2))'+f_1f_2f_3\sin\theta(\frac{f_3^2}{f_1^2f_2^2}
+\frac{f_1^2}{f_2^2f_3^2}+\frac{f_2^2}{f_1^2f_3^2})=0\nn \eea
This is satisfied given \p{fi'}. So indeed we have arrived at a
system of BPS equations for the four unknown functions.

To solve the BPS equations we first define a new radial variable
by $dt=f_1f_2f_3d\l$. In terms of this the equations become: \bea
\frac{d}{d\l}(\log f_i)&=&\frac{1}{2\cos \theta}\sum_i
f_i^2-\cos\theta f_i^2\nn \frac{d\theta}{d\l}&=&-\sin\theta \sum_i
f_i^2 \eea
Define $u_i=f_i^2\tan\theta$ then using the above we find:
\bea
\frac{d(u_i^{-1})}{d\l}&=&2\frac{\cos^2\theta}{\sin\theta}\nn
\frac{d\theta}{d\l}&=&-\cos\theta\sum_i u_i
\eea
Now define another radial variable by
$d\rho=(2\cos^2\theta/\sin\theta) d\l$. Now in terms of this we
can solve for $u_i$ and then for $\sin\theta$. We find that:
\bea\label{solution} u_i&=&\frac{1}{\rho-\a_i}\nn \sin\theta&=&(M
q(\rho,\a_i))^{-1/2} \eea where $q(\rho,\a_i)\equiv \prod_i
(\rho-\a_i)$.The $\a_i$ and $M$ are four arbitrary integration
constants. By rescaling the radial coordinate we find that the
solution takes the form: \bea\label{metricsolut}
ds^2&=&\frac{d\rho^2}{4\sqrt{q-M^2}}+\sum_i
\frac{\sqrt{q-M^2}}{\rho-\a_i}~g_i\nn
e^{2\Phi}&=&e^{2\Phi_0}(1-\frac{M^2}{q})\nn H&=&M(\sum_i
\frac{\omega_i}{(\rho-\a_i)^2}\wedge
d\rho-\frac{1}{2}\sum_i\frac{1}{\rho-\a_i}~\a) \eea In the limit
$M\rightarrow 0$ the torsion vanishes, the dilaton tends to a
constant and we recover the metric of \cite{Gibb2}. This is a
$G_2$ holonomy metric with a conical singularity for generic
values of the $\a_i$ but is regular when two of these constants
are equal. In this case one obtains the $G_2$ holonomy metric on
the $\mathbb{\RR}^3$ bundle over $\mathbb{CP}^2$ \cite{bs}. For
non-zero $M$ the torsion is non-vanishing and in the large $\rho$
limit the solution approaches the one in \cite{Gibb2}. In the
interior we see that the radial variable is constrained by
$\rho\geq\rho_0$ where $\rho_0$ is the solution of $q-M^2=0$. Note
that we always have $\rho_0\geq \a_i$. At $\rho=\rho_0$ the metric
is singular for all values of $\a_i$.
\par 
When $M=0$ the $G_2$ holonomy manifolds do
not have any compact associative three-cycles on which to wrap a
fivebrane, but they do have non-compact associative three-cycles. 
In the example of the $G_2$-holonomy metric on the
$\mathbb{\RR}^3$ bundle over $\mathbb{CP}^2$ there is 
a co-associative $\mathbb{CP}^2$ bolt, and the $\mathbb{\RR}^3$
fibres are non-compact and associative. It is thus natural
to interpret the solutions with $M\ne 0$ as describing
fivebranes wrapping such a
non-compact associative three-cycle, in the near horizon limit.
        
Finally, we point out that it should be very straightforward to
generalise the solutions in this section to co-homogeneity one
solutions where the principle orbits are $\mathbb{CP}^3$.
These include the $G_2$ holonomy metric on the 
$\mathbb{\RR}^3$ bundle over $S^4$ \cite{bs}.

\section{Recovering the solution of \cite{gr,GKMW}}

In this section we will use a similar procedure to recover the
solutions of \cite{gr,GKMW}. In so doing we will explicitly
demonstrate the $SU(3)$ structure of these solutions.

We first introduce frame one forms as in \cite{GKMW}:
\bea 
   &\nu^1&=d\theta \nn 
   &\nu^2&=\sin\theta d\phi\nn 
   &S^1&=\cos\phi\frac{\sigma^1}{2} - \sin\phi\frac{\sigma^2}{2}\nn 
   &S^2&=\sin\theta\frac{\sigma^3}{2} - 
      \cos\theta \left( \sin\phi\frac{\sigma^1}{2}
         +\cos\phi\frac{\sigma^2}{2} \right)\nn 
   &S^3&=-\cos\theta\frac{\sigma^3}{2} - 
      \sin\theta \left(\sin\phi\frac{\sigma^1}{2}
         +\cos\phi\frac{\sigma^2}{2} \right) \eea 
where $\theta,\phi$ are angles on a two-sphere and
$\sigma^a$ are the usual left invariant one forms on $SU(2)$
satisfying $d\sigma^a=\frac{1}{2}\e^{abc}\sigma^b\wedge\sigma^c$.
These satisfy the exterior algebra: \bea &d S^1&=2 S^2\wedge
S^3+\nu^2\wedge S^3+A\wedge S^2\nn
&d S^2&=2S^3\wedge S^1-\nu^1\wedge S^3-A\wedge S^1\nn
&d S^3&=2S^1\wedge S^2+\nu^1\wedge S^2-\nu^2\wedge S^1\nn
&d\nu^1&=0\nn &d\nu^2&=-A\wedge\nu^1 \eea where $A=\cos \theta
d\phi$. 
We introduce a frame: \bea &e^r&=a(r,x_7) dr\nn
&e^a&=b(r,x_7) S^a\nn &\tilde{e}^a&=c(r,x_7)(\nu^a+S^a)\nn
&e^3&=b(r,x_7) S^3\nn &e^7&=e^{\Phi}dx_7 \eea where $a=1,2$, and
make an ansatz for the $SU(3)$ invariant forms: 
\bea
&K&=e^7\nn
&\omega&=e^r\wedge e^3+e^1\wedge \tilde{e}^2-e^2\wedge \tilde{e}^1\nn
&\chi&=e^r\wedge(-e^1\wedge
e^2+\tilde{e}^1\wedge\tilde{e}^2)-e^3\wedge(e^1\wedge\tilde{e}^1+e^2\wedge\tilde{e}^2)\nn
&i_K*\chi&=e^r\wedge(e^1\wedge\tilde{e}^1+e^2\wedge\tilde{e}^2)-e^3\wedge(e^1\wedge
e^2-\tilde{e}^1\wedge \tilde{e}^2)
\eea corresponding to the metric: \bea ds^2=a^2
dr^2+b^2d\Omega_3^2+c^2\sum_{a=1,2}
(\nu^a+S^a)^2+e^{2\Phi}dx_7^2\eea where the orientation is taken
to be $\e^{r 3 1 \tilde{1}2\tilde{2} 7}=-1$.

The following identities are useful: \bea
&d\a&=-\frac{1}{2}d\beta=S^3\wedge \gamma\nn
&d\gamma&=2S^3\wedge\b \eea 
where 
\bea &\a&=S^1\wedge S^2\nn
&\b&=S^1\wedge \nu^2-S^2\wedge \nu^1+\nu^1\wedge \nu^2 \nn
&\gamma&=S^1\wedge \nu^1+S^2\wedge \nu^2 \eea Using these we can
write \bea
&\omega&=a b~dr\wedge S^3+b c~dS^3\\
&\chi&=a(c^2-b^2)dr\wedge \a+a c^2~ dr\wedge \b+\frac{c b^2}{2}d\b\\
&i_K *\chi&=a b c~ dr\wedge \gamma+b(c^2-b^2)S^3\wedge \a+\frac{b
c^2}{2}d\gamma \eea

Imposing the necessary and sufficient conditions discussed in
section 4.2 we find the following equations must be imposed: 
\bea\label{firstorder}
   \partial_{x_7}(e^{-\Phi}a b)&=&0\nn
   \partial_{x_7}(e^{-\Phi}b c)&=&0\nn
   \partial_{r}(e^{-\Phi}b c)&=&e^{-\Phi}a b\nn
   \partial_{r} b&=&\frac{a c}{b}
\eea 
and that the torsion is given by
\bea 
   H=F_1~ d\a+F_2~dr\wedge
\a+F_3~dr\wedge \b+F_4 ~dx_7\wedge \a+F_5~dx_7\wedge \b 
\eea 
where
\bea 
   &F_1&=-b^2 c e^{-\Phi} \partial_{x_7} \log(abc~e^{-2\Phi})\nn
   &F_2&=-a e^{-\Phi}(b^2\partial_{x_7}\log(b c^2
      e^{-2\Phi})-c^2\partial_{x_7}\log(b^3e^{-2\Phi})) \nn 
   &F_3&=a c^2e^{-\Phi}\partial_{x_7}\log(b^3e^{-2\Phi})\nn
   &F_4&=e^{\Phi}(\frac{b^2}{a}\partial_r \log(b c^2
      e^{-2\Phi})-\frac{c^2}{a}\partial_r\log(b^3e^{-2\Phi})
      -\frac{2b^2}{c}+\frac{2 c^3}{b^2}) \nn
   &F_5&=e^{\Phi}(-\frac{c^2}{a}\partial_r \log(b^3e^{-2\Phi})+\frac{2
      c^3}{b^2}) 
\eea 
Now imposing closure of $H$ we find that 
\bea
   \partial_r F_1+2F_3-F_2&=&0\nn
   \partial_{x_7}F_1+2F_5-F_4&=&0\nn
   \partial_{x_7}F_2-\partial_r F_4&=&0\nn
   \partial_{x_7}F_3-\partial_r F_5&=&0
\eea 
Using the first order equations (\ref{firstorder}) we find
that the above equations reduce to the single second order equation
\bea\label{secondorder} 
   a b e^{-\Phi}\partial_{x_7}^2 b+\partial_r(c e^{\Phi})=0 
\eea

Now (\ref{firstorder}) imply that $a b e^{-\Phi}=h(r)$ and by
choosing the radial coordinate appropriately we can set $h\equiv
1$. Then the rest of the equations determine the dilaton and $a,c$
in terms of $b$ via: 
\bea a^2&=&\frac{b}{r}\partial_r b\nn
c^2&=&a^2 r^2 \nn e^{\Phi}&=& a b \eea where $b$ satisfies the
second order non-linear pde: 
\bea\label{pde}
\frac{\partial^2}{\partial r ^2} b^3+3\frac{\partial^2}{\partial
{x_7}^2} b=0 \eea We recover the solution of \cite{GKMW} by
making a change of variables to $(z,\psi)$ such that :
\bea\label{newvariables} r&=&\sqrt{z}B(z)\sin \psi \nn
x_7&=&A(z)\cos \psi \eea where $A(z)$ and $B(z)$ satisfy 
\bea
A'(z)&=&B(z)\nn B'(z)&=&A(z)-\frac{B(z)}{2z} 
\eea 
so that 
\bea
   A(z)&=& \frac{z^{1/4}}{\sqrt{2}}(I_{-1/4}(z)+\mu~ K_{1/4}(z))\nn
   B(z)&=& \frac{z^{1/4}}{\sqrt{2}}(I_{3/4}(z)-\mu~ K_{3/4}(z)) 
\eea 
Here $\mu$ is an integration constant. The solution is then just
\bea 
   b^2=z
\eea 
Thus we have explicitly demonstrated the $SU(3)$ structure of the solution
found in \cite{gr,GKMW}. 

It seems a formidable challenge to find the general solution of 
\p{pde}. Let us just note that it easy to construct solutions
which do not depend on $x_7$. We then have
\bea\label{delocilized} 
b=(\l_1 r+\l_2)^{1/3} 
\eea 
These solutions might be interpreted as solutions corresponding
to wrapped NS fivebranes that are smeared over the $x^7$ direction.
Note in particular, that
the torsion is non vanishing for any choice of
the constants $\l_1,\l_2$ so that we do not recover the pure
geometry $CY_3\times S^1$, where $CY_3$ is the deformed conifold, as one
might have expected. The reason for this is simply that a more general 
ansatz for the $SU(3)$ structure is required. Enlarging our ansatz would
also allow one in principle to obtain more general wrapped NS fivebrane
solutions as well, but we expect that the pdes will be intractable without
further insight.

\section{Discussion}

We have analysed supersymmetric type II geometries of the form
$\RR^{1,5-d}\times M_{d+4}$ with non-trivial NS three-form flux and
dilaton, motivated by the fact that the near-horizon limits of wrapped
NS fivebranes geometries are of this type. In particular, we considered
the examples of seven-dimensional manifolds arising from branes
wrapped on associative or SLAG three-cycles. These geometries admit a
$G_2$ or an $SU(3)$ structure, respectively, of a specific type that
we determined. We also proved a converse result, namely that given
such a geometric structure then one obtains a supersymmetric solution
to the equations of motion. We used the converse result as a method to
construct solutions. Note that for both cases the group $G$ in the
$G$-structure is exactly the same as that of the underlying special
holonomy group of the manifold containing the supersymmetric cycle on
which fivebrane is wrapped. 

It is straightforward to extend the results for these specific
examples to the geometries arising when type II NS5-branes wrap other
supersymmetric cycles. In~\cite{GKMW} we analysed the holonomy of 
the connections $\nabla^\pm$ that would arise in each case, and a
summary appeared in table~1 of that reference. In $n$ dimensions,
in the cases where just one of the connections $\nabla^\pm$ has
special holonomy $G\subset SO(n)$, the geometry is specified by a
$G$-structure of a type that can be easily specified, by
following the discussion of section~2 (for related work see
\cite{strominger,Hull,IvPap,Iv3}). In the cases where both
$\nabla^\pm$ have special holonomy contained in $G'\subset SO(d)$
say, we find that the manifolds admit a $G\subset G'$-structure.
For both cases, one finds that the group $G$ of the $G$-structure
that appears in the final geometry is exactly the same as the
special holonomy group of the manifold, just as for the examples
explicitly discussed in this paper.

For example, $D=6$ geometries can arise when NS fivebranes wrap
K\"{a}hler two-cycles in Calabi-Yau three-folds or two-folds. In the
former case one of the connections $\nabla^\pm$ has special holonomy
$SU(3)$ while the other has general holonomy $SO(6)$. The resulting
geometry has an $SU(3)$-structure which was discussed in
\cite{strominger,Hull,IvPap}. Examples of this geometry were
presented in \cite{malnuntwo}. On the other hand when
NS-fivebranes wrap two-cycles in a Calabi-Yau two-fold we find
that both connections $\nabla^\pm$ have $SU(3)$-holonomy. In this
case the structure group of the six-manifold is in fact $SU(2)$.
This structure includes a product structure which allows one to
choose co-ordinates with a four-two split to the metric, but it
is not a direct product. Examples of this kind of geometry were
presented in \cite{gkmwone,zaf}.

Finally, it is also worth mentioning that much of the discussion
applies to type I supergravity. The action and supersymmetry
transformations are recorded in the Appendix. For this case there
is only a single connection with totally anti-symmetric torsion
$\nabla^+$ and so supersymmetry will just give rise to a single
$G$-structure. Consider for example the $D=7$ case. Since the
variation of the dilatino and gravitino for the type I theory are
the same as for $\e^+$ we deduce that the $G_2$ structure is
exactly the same as that discussed in section 3. In addition we
need to ensure the vanishing of the supersymmetry variation of the
gaugino
\be 
   F_{mn}\gamma^{mn}\e=0 
\ee 
This implies, following~\cite{GN} that $F$ must satisfy the $G_2$
instanton equation $F_{mn}=\frac{1}{2}{*\phi}_{mn}{}^{pq}F_{pq}$
\textit{i.e.} the two-form $F$ is in the {\bf 14} in the decomposition
\p{twoforms}. This is the type of geometry dictated by supersymmetry
that would arise when type I fivebranes wrap associative three-cycles
and also SLAG three-cycles. To obtain a solution to the equations of
motion for type I supergravity we have to solve
%\footnote{Of course in
%  string theory one could more generally solve $dH=2\alpha'\Tr(F\wedge F
%  - R\wedge R)$. In terms of branes this leads to the difference between
%  wrapped symmetric and gauge fivebrane solutions~\cite{HS}.} 
$dH=2\alpha'\Tr F\wedge F$. Using the integrability conditions given
in the appendix it is clear that these conditions are also sufficient
to obtain a supersymmetric solution to the equations of motion, by
generalising the argument in section~2. As yet there are no known
solutions of this kind with non-vanishing $F$. 
Such solutions would have the geometry naturally
expected for type I or heterotic ``gauge''
fivebranes \cite{hs} wrapping associative three-cycles.
For the case of type I fivebranes wrapping SLAG three-cycles
the interesting
possibility arises that there will in fact be an $SU(3)$ structure,
despite the fact that it is not dictated by supersymmetry alone.  

It is interesting to note that the type II solutions give rise
to type I solutions with $F=0$. These correspond to
type I or heterotic ``neutral'' fivebranes \cite{hs} wrapping associative
three-cycles. The type II solutions corresponding to fivebranes
wrapping SLAG three-cycles thus give rise to type I
solutions describing type I fivebranes wrapping SLAG three-cycles
that have an $SU(3)$ structure. This is some evidence that this
will also occur for wrapped gauge-fivebranes.

In type I or heterotic string theory anomaly cancellation implies that the
Bianchi identity is modified by higher order corrections in $\alpha'$.
To leading order this is most informatively written as
$dH=2\alpha'\Tr[F\wedge F  - R(\Omega^-)\wedge R(\Omega^-)]$ where
$\Omega^-=\omega-H/2$ \cite{deroo}. This should be viewed as
implicitly defining $H$. One can ask whether one can solve this
for wrapped branes by identifying $A$ with $\Omega^-$ as this would be
the analogue of ``symmetric fivebranes'' \cite{hs}. For supersymmetric 
fivebranes wrapping associative three cycles, only $\Omega^+=\omega+H/2$
has holonomy contained in $G_2$ and hence identiying $A$ with $\Omega^-$
would not be supersymmetric. Interestingly, for 
supersymmetric fivebranes wrapping SLAG three-cycles both
$\Omega^\pm$ have holonomy contained in $G_2$ and hence one can 
obtain supersymmetric solutions for these cases. More explicitly,
the solution constructed in \cite{GKMW} automatically gives a solution
of the heterotic or type I string theory
with non-vanishing gauge-fields
if we simply identify $A=\Omega^-$.
This argument equally applies to the solutions found in \cite{gkmwone,zaf}
corresponding
to fivebranes wrapping two-cycles in Calabi-Yau two-folds. For the type
II theory these are holographically dual to a slice of the Coulomb
branch of pure ${\cal N}=2$ super-Yang-Mills theory \cite{gkmwone}. 
The corresponding type
I solution has half the supersymmetry and so should holographically
encode information about the ${\cal N}=1$ gauge theories arising on type
I or heterotic fivebranes wrapping two-cycles in Calabi-Yau two-folds.
It would be interesting to study this further.

%%%%%%%%%%%%%%%%%%%%%%%%%%%%%%%%%%%%%%%%%%%%%%%%%%%%%%%%%%%%%%%%%%%%%%%%%%%

\subsection*{Acknowledgements}

We would like to thank Gary Gibbons and Chris Hull for helpful discussions. All
authors are supported in part by PPARC through SPG $\#$613. DW also
thanks the Royal Society for support. 

%%%%%%%%%%%%%%%%%%%%%%%%%%%%%%%%%%%%%%%%%%%%%%%%%%%%%%%%%%%%%%%%%%%%%%%%%%%

\section{Appendix}

We will derive the integrability conditions in the context of
type I SUGRA. The bosonic fields are the same as the NS sector of
the type II supergravity supplemented by a gauge field in the
adjoint of some gauge group, with field strength $F$. For gauge
group $SO(32)$ or $E_8\times E_8$ this is part of the low-energy
effective action of type I or heterotic string theory. The action
is given by: \be S=\frac{1}{2\kappa^2}\int d^{10} x \sqrt {-g}
e^{-2\Phi} \left(R+4(\nabla\Phi)^2-\frac{1}{12}H^2 - \alpha' \Tr
F^2\right) \ee with \be dH=2\alpha' \Tr F\wedge F \ee The equations
of motion are given by \bea
R_{\mu\nu}-\frac{1}{4}H_{\mu\rho\sigma}H_{\nu}{}^{\rho\sigma}+2\nabla_\mu
\nabla_\nu\Phi -2 \alpha' \Tr F_\mu{}^\rho F_{\nu\rho}=0\nn
\nabla^2\Phi-2(\nabla\Phi)^2+\frac{1}{12}H_{\mu\nu\rho}H^{\mu\nu\rho}
+\frac{\alpha'}{2}\Tr F_{\mu\nu}F^{\mu\nu}=0\nn
\nabla_\mu(e^{-2\Phi}H^{\mu\nu\rho})=0\nn
2e^{2\Phi}D^\mu(e^{-2\Phi}F_{\mu\nu})-
F^{\rho\sigma}H_{\rho\sigma\nu}=0 \eea

Supersymmetric configurations have vanishing variation of the
gravitino, dilatino and gaugino: \bea \delta \psi_\mu &\sim&
\nabla_\mu \epsilon
+\frac{1}{8}H_{\mu\nu\rho}\Gamma^{\nu\rho}\epsilon = 0\nn \delta
\lambda &\sim&  \Gamma^\mu\partial_\mu \Phi\epsilon +
\frac{1}{12}H_{\mu\nu\rho}
\Gamma^{\mu\nu\rho}\epsilon = 0 \nonumber\\
\delta\chi&\sim&F_{\mu\nu}\Gamma^{\mu\nu}\epsilon=0 \label{gravitino}
\eea 
where $\epsilon$ is a Majorana-Weyl spinor of $Spin(1,9)$.
Note that the first two conditions are half of the conditions
arising in the type II theories. We now deduce some consequences 
of the integrability
conditions of these equations.

First, take the covariant derivative of the variation of the
gravitino and antisymmetrise, to get \be
R_{\mu\nu\s_1\s_2}\Gamma^{\s_1\s_2}\e=
-\nabla_{[\mu}H_{\nu]\s_1\s_2}\Gamma^{\s_1\s_2}\epsilon
-\frac{1}{2}H_{[\mu}{}^{\s_1}{}_{|\rho|}
H_{\nu]}{}^{\rho\s_2}\epsilon \ee Next multiply this expression
by $\Gamma^\nu$ and use a Bianchi identity to obtain an
expression for $R_{\mu\nu}\Gamma^\nu\epsilon$. Then use
$H_{\mu\nu\rho}\Gamma^{\nu\rho}$ times the dilatino variation,
the covariant derivative of the dilatino as well as
$F_{\mu\nu}\Gamma^\nu$ times the variation of the gaugino to get
\begin{multline}\label{gravint}
(R_{\mu\nu}-\frac{1}{4}H_{\mu\rho\sigma}H_{\nu}{}^{\rho\sigma}+2\nabla_\mu
\nabla_\nu\Phi -2 \alpha' \Tr F_\mu{}^\rho F_{\nu\rho})\Gamma^\nu\epsilon
=\\
\frac{1}{12}(dH-2\alpha' \Tr F\wedge F)_{\mu\nu\rho\sigma}
\Gamma^{\nu\rho\sigma}\epsilon
+\frac{1}{2}e^{2\Phi}\nabla^\rho(e^{-2\Phi}H_{\rho\mu\nu}\Gamma^\nu)\epsilon
\end{multline}

Similar manipulations on $\Gamma^\mu\nabla_\mu$ acting on the
variation of the dilatino implies
\begin{multline}
(\nabla^2\Phi-2(\nabla\Phi)^2+\frac{1}{12}H_{\mu\nu\rho}H^{\mu\nu\rho}
+\frac{\alpha'}{2}F_{\mu\nu}F^{\mu\nu})\epsilon=\\
-\frac{1}{48}(dH-2\alpha'F\wedge
F)_{\mu\nu\rho\sigma}\Gamma^{\mu\nu\rho\sigma} \epsilon
-\frac{1}{4}e^{2\Phi}\nabla^\mu(e^{-2\Phi}H_{\mu\nu\rho})\Gamma^{\nu\rho}
\epsilon
\end{multline}
while $\Gamma^\mu\nabla_\mu$ acting on the variation of the
gaugino yields \be\label{fint}
(2e^{2\Phi}\nabla^\mu(e^{-2\Phi}F_{\mu\nu})-F^{\rho\sigma}H_{\rho\sigma\nu})
\Gamma^\nu\epsilon=3DF_{\mu\nu\rho}\Gamma^{\mu\nu\rho}\epsilon \ee

We next note that if the Bianchi identities for $H$ and $F$
satisfied as well as the $H$ equation of motion then we deduce
the dilaton equation of motion. The equation \p{fint} is of the
form $A_\mu \Gamma^\mu\epsilon=0$ which implies $A^\mu A_\mu=0$.
Similarly \p{gravint} is of the form $B_{\mu\nu}\Gamma^\nu=0$
which implies  $B_{\mu\nu}B^{\mu\nu}=0$. If we assume that we
have a solution of the form $\RR^{1,9-n}\times M_n$ then we can
deduce that $A_{m}=B_{mn}=0$ which give the gauge and Einstein
equations of motion. In other words the Killing spinor equations,
combined with the Bianchi identities for $H$ and $F$, plus the
$H$ equations of motion imply all equations of motion are
satisfied.

\end{document}